\documentclass[a4paper,11pt]{article}
\pdfoutput=1 

\usepackage{jheppub} 
\usepackage{graphicx}  
\usepackage{dcolumn}   
\usepackage{bm}        
\usepackage{amssymb}   
\usepackage{amsmath}
\usepackage{color}
\usepackage{bbold}
\usepackage{slashed}
\usepackage{epsfig}
\usepackage{epstopdf}
\usepackage[usenames,dvipsnames]{xcolor}
\usepackage{comment}
\usepackage[T1]{fontenc} 
\usepackage{color}
\begin{document}

\hspace{5.2in} \mbox{NSF-KITP-15-152}
\title{{\bf Quantum gravity of Kerr-Schild spacetimes and the logarithmic correction to Schwarzschild black hole entropy}}

\author{Basem Kamal El-Menoufi}

\affiliation{Department of Physics,
University of Massachusetts\\
Amherst, MA  01003, USA}

\emailAdd{bmahmoud@physics.umass.edu}

\abstract{In the context of effective field theory, we consider quantum gravity with minimally coupled massless particles. Fixing the background geometry to be of the Kerr-Schild type, we fully determine the one-loop effective action of the theory whose finite {\em non-local} part is induced by the long-distance portion of quantum loops. This is accomplished using the non-local expansion of the heat kernel in addition to a {\em non-linear completion} technique through which the effective action is expanded in gravitational curvatures. Via Euclidean methods, we identify a logarithmic correction to the Bekenstein-Hawking entropy of Schwarzschild black hole. Using dimensional transmutation the result is shown to exhibit an interesting interplay between the UV and IR properties of quantum gravity.}
\maketitle
\flushbottom

\section{Introduction}

General relativity is a well-behaved quantum theory at low energies \cite{JohnPRL,JohnEFT}. Treated as an effective field theory (EFT), quantum predictions can systematically be quantified. The clear separation of scales provided by the EFT framework enables the extraction of the leading quantum effects. The latter are precisely due to the low-energy portion of the theory which is dictated by the symmetries of general relativity. On the other hand, the unknown high-energy physics is manifested only in the Wilson coefficients of the most general Lagrangian. All observables are then expressed in terms of the low energy constants, which are experimentally measured. As an EFT, the theory is renormalizable order by order in the counting parameter, i.e. $E/M_\text{P}$, which makes it fully predictive.

Massless particles can propagate over long distances. The quantum fluctuations of massless excitations offer a unique feature in field theory; non-locality. For example, the non-analytic portion of scattering amplitudes is due to the low energy propagation of massless particles. Using EFT techniques, Donoghue and collaborators determined the leading long-distance modification to the Newtonian potential \cite{JohnPRL,JohnEFT,Johnnewton}. More generally, this class of quantum corrections establish a set of low-energy theorems of quantum gravity \cite{Johntheorem}. Apart from scattering amplitudes, previous investigations focused primarily on the regime of weak gravity where gravitons propagate through flat space. For instance, quantum corrections to various black hole geometries in the asymptotic region were computed in \cite{Johnschwarz}. 

It is very natural then to pose the following question: What is the full structure of the loop-induced modifications to general relativity? In order to treat the non-linear regime of gravity, we clearly need to quantify these infrared corrections in curved spacetimes. Here, the technical aspect concerns the construction and properties of non-local effective actions. These are somewhat easy to understand in Minkowski space \cite{Basemtrace} but become quite complicated when considered in curved space \cite{Basem,BasemQED,Barvinskyreport,Barvinsky83,Barvinsky87,Barvinsky90,Barvinsky94,Avramidi}. The non-local corrections provide a {\em quantum memory} and could become appreciable even below the Planck scale \cite{Basem}. For example, the analysis presented in \cite{Basem} hints at the possible avoidance of cosmological singularities\footnote{See also a host of papers \cite{Espriu,Cabrer,Woodard1,Woodard2,Woodard3,Woodard4,Calmet,Maggiore} that explore the phenomenology of non-local models.}.
   
On a different front, the startling discovery that a black hole is a thermodynamic system endowed with entropy stands out as a remarkable achievement of twentieth century physics. A complete understanding of the Bekenstein-Hawking (BH) area law \cite{Bekenstein,Hawkingcreation} is believed by many to be our window to learn profound lessons about quantum gravity. There exist plenty of {\em macroscopic} derivations of the BH entropy using different approaches that we briefly discuss below. Nevertheless, the conundrum we face concerns the statistical or {\em microscopic} description of black hole entropy. There has been partial success to address this question in string theory \cite{Vafa}, holography \cite{Strominger} and quantum geometry \cite{Rovelli} but we are still far from a definitive answer\footnote{We only include a restricted list of references since microscopic derivations lie beyond the scope of this work.}. 

It is well known that the BH area law does not hold in more general theories of gravity \cite{Jacobsonprd,Jacobsonprl}. In this light it is crucial to study quantum corrections to Einstein gravity and their corresponding effect on the area law. Thus, even on the macroscopic side it is quite possible to gain new insights about quantum gravity. One might nevertheless be tempted to think that an exact knowledge of these deviations requires a UV completion of gravity. This is certainly not the case if the corrections emerge from the infrared limit of quantum loops of massless particles. As described above, these {\em parameter-free} corrections are genuine predictions of quantum gravity. Once known, they furnish a test laboratory for any proposed UV completion.

In this paper, we adopt the EFT framework to study quantum gravity\footnote{See \cite{Donoghueeftreview,Burgess} for detailed reviews.} with free massless minimally coupled (MMC) matter fields in Kerr-Schild (KS) spacetimes. For KS spacetimes, there exist coordinates such that the spacetime metric reads\footnote{Throughout this paper, we assume a vanishing cosmological constant.}
\begin{align}\label{ksansatz}
g^{\text{KS}}_{\mu\nu} = \eta_{\mu\nu} - k_\mu k_\nu 
\end{align}
with $k^\mu$ - the KS vector - being a null vector field. It is a remarkable fact that black holes in vacuum Einstein gravity are of the KS type. In particular, the Kerr solution was originally found using the KS ansatz, thanks to the extreme reduction in complexity provided by the formalism \cite{Kerr,Schild,Debney}. Although we do not intend to review the formalism at length\footnote{The interested reader could consult \cite{Chand,Stephani} for thorough accounts.}, a short version is provided in appendix \ref{KS}, where we show how one can obtain both the Schwarzschild and Kerr solutions starting from the KS ansatz. 

We have two goals in mind for the present paper: 
\begin{itemize}
\item To address some of the subtleties associated with the construction of non-local actions in curved spacetimes. Previous studies \cite{Basemtrace,BasemQED,Basem,Barvinskyreport,Barvinsky83,Barvinsky87,Barvinsky90,Barvinsky94,Avramidi} have focused on obtaining results appropriate for a generic metric. Albeit robust, the results are complicated for an arbitrary geometry and some questions remain unanswered in regard to the nature of the so called {\em form factors}. It is not clear whether the available results provide the best pathway to explore the phenomenology.

The special form of the KS metric enables us to {\em exactly} resolve the heat kernel for various operators. Hence, we can probe the structure of non-local actions in a non-trivial context. In spite of being special, the KS class contains black holes which are phenomenologically the most relevant. Our results pave the way to interesting further progress in the quantum physics of black holes.
 
\item To compute the logarithmic correction to the Schwarzschild black hole entropy. The non-analytic dependence on the horizon area hints that the underlying action is non-local. The effective action can readily be used to identify the logarithmic correction by constructing the Euclidean partition function. Moreover, knowledge of the partition function is a precursor to explore quantum aspects of black hole thermodynamics. We posit a few interesting questions in section \ref{sect5}.
\end{itemize}

A quick review of the literature regarding the mentioned goals is in place. First, a significant amount of work has been undertaken to uncover the structure of non-localities in gravitational effective actions, see \cite{Basemtrace,BasemQED,Basem,Barvinskyreport,Barvinsky83,Barvinsky87,Barvinsky90,Barvinsky94,Avramidi} and references therein. Results are customarily displayed as an expansion in gravitational curvatures. Nevertheless, this expansion is quite different from {\em local} Lagrangians familiar in (non)-renormalizable quantum field theories. For instance, the effective Lagrangian of quantum gravity is arranged according to the energy or derivative expansion and only local polynomials of curvature invariants appear. This is the typical story when one integrates out a heavy field from the path integral of the theory. On the other hand, quantum loops of massless fields yield a non-local effective theory. The so called form factors are fundamental objects in the non-local expansion and the covariance properties of the latter were scrutinized in \cite{BasemQED}. 

One great advantage of fixing the background geometry to have the KS form is an unambiguous definition of the form factors. In this case, the results turns out to be much simpler than those which exist in the literature \cite{Basemtrace,BasemQED,Basem,Barvinskyreport,Barvinsky83,Barvinsky87,Barvinsky90,Barvinsky94,Avramidi}. In addition, the KS form of the metric allows for a transparent analysis of the curvature expansion, which we shall review in section \ref{sect3}. The nature of the non-local expansion becomes manifest, which provides invaluable clues for future endeavors.

Moving to the second goal where a decent amount of work has been done as well. Fursaev, to the best we know, provided the first hint about the logarithmic correction in \cite{Fursaev} using the conical singularity method. Recently, Sen and collaborators used Euclidean methods to uncover the logarithmic correction for both extremal \cite{Gupta1,Gupta2,Sen1} as well as non-extremal \cite{Sen2} black holes. When available, the results remarkably agree with microscopic results in the extremal case. Carlip employed Cardy's formula, which counts states in 2$\text{d}$ conformal field theory, to find the logarithmic correction to the BH entropy \cite{Carlip}. The authors of \cite{Suneeta} computed the exact partition function of the BTZ black hole to uncover the logarithmic correction. Banerjee and collaborators used the tunneling approach to identify corrections to Hawking temperature which then yield a logarithmic correction in the entropy of various black holes \cite{Banerjee1,Banerjee2}. Other authors used the anomaly-induced action, i.e. Riegert action, to compute the same correction this time via Wald's Noether charge formalism \cite{Aros}. The authors of \cite{Cai} obtained exact black hole solutions to the semi-classical Einstein equations including the conformal anomaly. A direct computation revealed a logarithmic correction to the BH entropy. Finally, the logarithmic correction was also found based on the quantum geometry program \cite{Kaul}. 

Now we summarize our results. Our starting point is the EFT action
\begin{align}
\mathcal{S} = \mathcal{S}_{\text{GEFT}} + \mathcal{S}_{\text{matter}} \ \ .
\end{align} 
The gravitational effective action is 
\begin{align}\label{geftaction}
\mathcal{S}_{\text{GEFT}} = \int d^{\text{d}}x \, \sqrt{g} \, \left( \frac{M_{\text{P}}^2}{2} \, R + + c_1\, R^2 + c_2 \, R_{\mu\nu} R^{\mu\nu} + c_3\, R_{\mu\nu\alpha\beta} R^{\mu\nu\alpha\beta} + c_4\, \nabla^2 R \right)
\end{align}
where only operators containing up to four derivatives are included. Notice here that the above is not usually how the action is displayed \cite{JohnPRL,JohnEFT}. The last term is customarily omitted because it is a total derivative and does not contribute to the Feynman rules while the Riemann piece is omitted via an implicit use of the Gauss-Bonnet identity. We shall see below that we need to keep all the terms in order to carry out the renormalization program. The second portion $\mathcal{S}_{\text{matter}}$ describes free MMC matter fields of spin $0,1/2,1$. The constants $(c_1,c_2,c_3,c_4)$ are the {\em bare} Wilson coefficients\footnote{As per usual, the bare constants remain dimensionless in $\text{d}$ dimensions.} and the dimensionality of spacetime is extended in order to employ dimensional regularization, i.e. $\text{d} = 4 -2\epsilon$. The one-loop effective action is evaluated fixing the background geometry to be a KS spacetime. Upon integrating out the matter degrees of freedom and graviton fluctuations at the one-loop level\footnote{The one-loop graviton fluctuations arise solely from the Einstein-Hilbert action. There is indeed a contribution from the $\mathcal{O}(\partial^4)$ pieces but these are suppressed by the Planck mass. To be consistent with the power counting of the EFT, these are included only when one considers the two-loop effective action.}, we obtain
\begin{align}
\Gamma[\bar{g}] = \Gamma_{\text{local}} + \Gamma_{\text{ln}}
\end{align}
where the renormalized action now reads
\begin{align}\label{localactionintro}
\Gamma_{\text{local}}[\bar{g}] = \int d^4x \, \left( \frac{M_{\text{P}}^2}{2} \, R + c^r_1(\mu) \, R^2 + c^r_2(\mu) \, R_{\mu\nu} R^{\mu\nu} + c^r_3(\mu) \, R_{\mu\nu\alpha\beta} R^{\mu\nu\alpha\beta} + c^r_4(\mu)\, \nabla^2 R \right) \ \ .
\end{align}
Here, $\bar{g}$ is the background metric that takes the form in eq. (\ref{ksansatz}) and $\mu$ is the scale of dimensional regularization. Notice in particular that Newton's constant does {\em not} get renormalized because the divergences arising from massless loops are proportional to the quadratic invariants. Of utmost importance is the finite pieces that exhibit a logarithmic non-locality
\begin{align}\label{logactionintro}
\nonumber
\Gamma_{\text{ln}}[\bar{g}] = - \int d^4x \, &\bigg(\alpha\, R \ln\left(\frac{\Box}{\mu^2}\right) R + \beta\, R_{\mu\nu} \ln\left(\frac{\Box}{\mu^2}\right) R^{\mu\nu} \\
&+ \gamma\, R_{\mu\nu\alpha\beta} \ln\left(\frac{\Box}{\mu^2}\right) R^{\mu\nu\alpha\beta} + \Theta\, \ln\left(\frac{\Box}{\mu^2}\right) \Box R \bigg)
\end{align} 
where $\Box = \eta^{\mu\nu} \partial_\mu \partial_\nu$. The different coefficients depend on the particle species and are listed in table \ref{tablecoeff}.

Focusing on the Schwarzschild solution, we use the effective action to construct the partition function. From the latter, the entropy is determined and our main result reads
\begin{align}\label{entropyintro}
S_{\text{bh}} = S_{\text{BH}} + 64\pi^2 \left( c^r_3(\mu) + \Xi\, \ln \left(\mu^2 \mathcal{A} \right)\right) \ \ .
\end{align}  
Here $S_{\text{BH}}=\mathcal{A}/4G$ is the BH entropy and $\mathcal{A}=16 \pi (GM)^2$ is the horizon area. The constant $\Xi$ sums up the contributions from all the massless particles in the theory and reads
\begin{align}\label{sumparticles}
\Xi = \frac{1}{11520\pi^2} \left(2 N_s + 7 N_f - 26 N_V + 424 \right)
\end{align}
where we allowed for variable number of particles. The logarithmic dependence on the horizon area and the associated coefficient is in exact agreement with \cite{Fursaev,Sen2,Banerjee1,Aros} albeit using different approaches than ours. Furthermore, eq. (\ref{entropyintro}) contains a subtle feature: the entropy is manifestly renormalization-group (RG) invariant. The demonstration of this property is made clear in section \ref{sect4}. In fact, this feature is mandatory if black hole entropy is to be identified as a physical quantity. We can further employ dimensional transmutation to rewrite eq. (\ref{entropyintro}) as
\begin{align}
S_{\text{bh}} = S_{\text{BH}} + 64 \pi^2 \, \Xi \, \ln \left(\frac{\mathcal{A}}{\mathcal{A}_{\text{QG}}}\right)
\end{align}
where $\mathcal{A}_{\text{QG}}$ corresponds to a length (energy) scale {\em uniquely} set by the full theory, i.e. the UV completion of quantum gravity. As we shall discuss further below, the result uncovers an intricate connection between the UV and IR properties of quantum gravity. More comments about the content of the result are reserved to section \ref{seccomments}.

The plan of the paper is as follows. We commence in section \ref{sect1} by developing a set of Feynman-like rules to resolve the heat kernel for the d' Alembertian operator in KS spectimes. The Einstein equations are solved with the KS ansatz in appendix \ref{KS} while the non-local expansion of the heat kernel is described in appendix \ref{hkernel}. In section \ref{sect2} the curvature expansion is introduced and the technique of non-linear completion is used to express the heat kernel trace in the desired form. We then move in section \ref{sect3} to find the effective action by integrating over proper time. There, we uncover what we would like to call a {\em UV-IR correspondence}. Among other things, this correspondence allows us to extend the results to matter fields of various spins and gravitons. This is acheived knowing only the divergences of the theory. In section \ref{sect4} the partition function is determined using the effective action. The behavior of the partition function under a global scale transformation provides an elegant pathway to extract the logarithmic correction to the BH entropy. We discuss possible future directions in section \ref{sect5}. In appendix \ref{usefuliden} we collect useful formulas used throughout the paper. 

\section{The heat kernel for the covariant d' Alembertian}\label{sect1}

In this section, we commence by resolving the heat kernel of the d' Alembertian operator. Knowing the latter enables a straightforward determination of the effective action which results from integrating out a massless free scalar. One can otherwise directly compute the effective action via Feynman graphs \cite{Basemtrace,Basem} but we choose to work with the heat kernel for reasons that we shall spell out below. The basic definitions and properties of the heat kernel are given in appendix \ref{hkdef}. Now we restrict our consideration to KS spacetimes of the form displayed in eq. (\ref{ksansatz}). An immediate consequence of the null property of the KS vector is the set of relations
\begin{align}\label{kssimple}
\sqrt{g} = 1, \quad g^{\mu\nu} = \eta^{\mu\nu} + \lambda\, k_\mu k_\nu, \quad g^{\mu\nu} k_\mu k_\nu = \eta^{\mu\nu} k_\mu k_\nu = 0 
\end{align}
where the Minkowski metric is expressed in standard coordinates. Here $\lambda$ is a trivial counting parameter which is set to unity at the end of the computation. In order to treat operators with no associated mass scale, we use the non-local expansion of the heat kernel developed by Barvinsky, Vilkovisky and collaborators \cite{Barvinsky83,Barvinskyreport,Barvinsky87,Barvinsky90}. For the convenience of the reader, we provide an essential review of the formalism in appendix \ref{hkexpansion}. 

We seek an expansion of the heat kernel in powers of $\lambda$. Let us quote the d' Alembertian operator as it acts on a scalar density of weight $1/2$
\begin{align}\label{dalemw12}
\nabla^2 \Psi = \frac{1}{\sqrt[4]{g}}\, \partial_\mu \left(\sqrt{g} g^{\mu\nu} \partial_\nu\right) \frac{1}{\sqrt[4]{g}} \, \Psi\ \ .
\end{align}
The KS form of the metric drastically simplifies the structure of the operator
\begin{align}\label{dalemKS}
\nabla^2 \Psi = \left(\partial^2 + \lambda\, k^\mu k^\nu \partial_\mu \partial_\nu + \frac{\lambda}{2} \, \partial_\mu (k^\mu k^\nu) \partial_\nu + \frac{\lambda}{2}\, \partial_\nu (k^\mu k^\nu) \partial_\mu \right) \Psi \ \ .
\end{align}
It is important to pause at this stage and comment on the above result. Let us imagine that we aim to study the same operator on a generic background spacetime. The conventional treatment is to expand the metric around flat space as $g_{\mu\nu} = \eta_{\mu\nu} + \text{H}_{\mu\nu}$ and proceed to evaluate the heat kernel in powers of the {\em external} classical field $\text{H}_{\mu\nu}$. Both the inverse metric and metric determinant are expanded accordingly and the result is an infinite series in $\text{H}_{\mu\nu}$. Consequently the d' Alembertian operator contains arbitrarily high powers of the external field. On the contrary, there is an immediate truncation for KS spacetimes as evident from eq. (\ref{dalemKS}). More comments about similar simplifications are made as we go along. 

In the notation of appendix \ref{hkexpansion}, we identify the interaction term
\begin{align}\label{ksint}
V = \lambda \left( k^\mu k^\nu \partial_\mu \partial_\nu + \frac{1}{2} \partial_\mu (k^\mu k^\nu) \partial_\nu + \frac{1}{2} \partial_\nu (k^\mu k^\nu) \partial_\mu \right) \ \ .
\end{align}
For later convenience, we define the following tensor
\begin{align}
K_{\mu\nu} \equiv k_\mu k_\nu \ \ .
\end{align}
We seek an expansion of the heat kernel {\em trace} in powers of $\lambda$. Using eqs. (\ref{traceheatmat}) and (\ref{tevolutionfinal}) one can easily introduce Fourier transforms to derive a set of Feynman-like rules which read:
\begin{figure}\label{frules}
\centerline{\includegraphics[width=4cm,height=4cm]{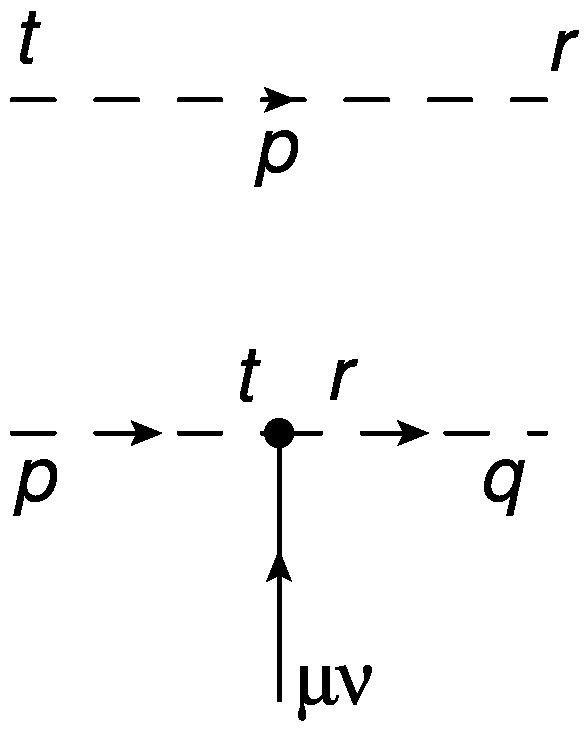}
\put(10,100){$e^{(t+r)p^2}$}
\put(10,45){$p_{(\mu}q_{\nu)}\, \delta(t+r)$}}
\caption{Feynman-like rules for the heat kernel trace. The solid line corresponds to an insertion of the external tensor field $K_{\mu\nu}$ which carries a power of $\lambda$.}
\end{figure}

\begin{itemize}
\item The rule for the vertex and propagator are given in the figure \ref{frules}.
\item The internal propagator that carries the loop momentum gets an extra factor of 1 in the exponent\footnote{This is due to the flat space kernel that appears convoluted in eq. (\ref{traceheatmat}).}.
\item Add a factor of $s$ in the exponent of all propagators.
\item Impose momentum conservation at each vertex.
\item Integrate over the loop momentum and proper-time\footnote{Here, we mean the integration variable in the exponent of eq. (\ref{tevolutionfinal}).}. 
\end{itemize}

\begin{figure}\label{expansion}
\centerline{\includegraphics[width=8cm,height=3cm]{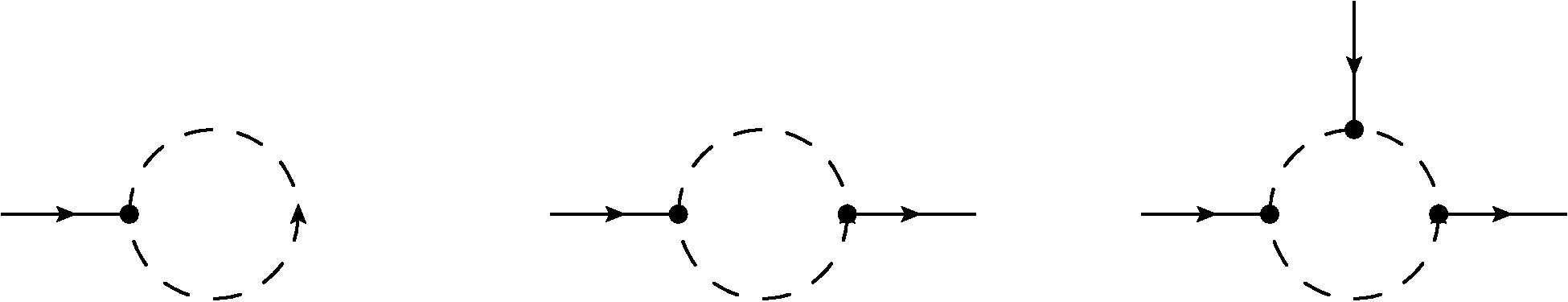}}
\caption{The diagramatic expansion of the heat kernel trace.}
\end{figure}

From the Feynman-like rules, we easily develop a diagramatic expansion as shown in figure \ref{expansion}. Here, a great simplification emerges thanks to the KS form of the metric: there is a single diagram in the expansion at each order in $\lambda$. On the contrary, for a generic background the number of diagrams proliferate as we go to higher orders in the expansion.

\subsection{Lowest order}\label{lowestorder}

Let us compute the first diagram in figure \ref{expansion}. Applying the rules given above, we find
\begin{align}
\overset{\scriptscriptstyle{(1)}}{\mathcal{H}}(s) = s K^{\mu\nu}_0 \int \frac{d^{\text{d}}l}{(2\pi)^{\text{d}}}\,  l_\mu l_\nu \, e^{sl^2}
\end{align}
where the subscript on the background field denotes its momentum, i.e. $ K^{\mu\nu}_0 \equiv K_{\mu\nu}(0)$. Using the tensor integrals given in appendix \ref{tensorint} we find 
\begin{align}
\overset{\scriptscriptstyle{(1)}}{\mathcal{H}}(s) = - \frac{i}{2(4\pi s)^{{\text{d}}/2}} K^{\mu\nu}_0 \eta_{\mu\nu}\ \ .
\end{align}
By construction the KS vector is null with respect to the Minkowski metric, and thus
\begin{align}\label{htrace1}
\overset{\scriptscriptstyle{(1)}}{\mathcal{H}}(s) = 0 \ \ .
\end{align}
This is the trace of the heat kernel to lowest order in $\lambda$ and the result is exact. Nevertheless, we shall see in the next section that we need to compute the heat kernel in the coincidence limit rather than the trace. This is necessary in order to carry out the non-linear completion procedure that we explain in the next section.

\subsection{Next-to-leading order}
At $\mathcal{O}(\lambda^2)$ we encounter the second diagram in figure \ref{expansion}. We display the steps in some detail to elucidate the construction. Straightforward application of the rules yields
\begin{align}\label{secordertrace}
\nonumber
\overset{\scriptscriptstyle{(2)}}{\mathcal{H}}(s) = s^2 \int \frac{d^{\text{d}}p}{(2\pi)^{\text{d}}}\, K^{\mu\nu}_p K^{\alpha\beta}_{-p} \int_0^1 dt_1 \int_0^{t_1} dt_2 \int &\frac{d^{\text{d}}p}{(2\pi)^{\text{d}}}\, V_{\mu\nu}(l,p) V_{\alpha\beta}(l,p)\\
&e^{s\left((1-t_1+t_2)l^2 + (t_1-t_2)(l+p)^2 \right)}
\end{align}
where
\begin{align}
V_{\mu\nu}(l,p) = l_\mu l_\nu + l_{(\mu} p_{\nu)} \ \ .
\end{align}
We first need to put the exponent in eq. (\ref{secordertrace}) in quadratic form. In particular, this enables dropping odd powers of the loop momentum. This is accomplished via shifting the loop momentum by sending $l \to l + (t_1-t_2) p$. If we moreover perform the tensor integrals using appendix \ref{tensorint} we find 
\begin{align}
\nonumber
\overset{\scriptscriptstyle{(2)}}{\mathcal{H}}(s) = s^2 & \int \frac{d^{\text{d}}p}{(2\pi)^{\text{d}}} K^{\mu\nu}_p K^{\alpha\beta}_{-p} \int_0^1 dt_1 \int_0^{t_1} dt_2 e^{\sigma(1-\sigma)sp^2} \\\nonumber
&\big[J_{\mu\nu\alpha\beta} - \sigma(1-\sigma) \left(J_{\mu\nu}p_\alpha p_\beta + J_{\alpha\beta} p_\mu p_\nu \right) + \frac{1}{4} (1-2\sigma)^2 \big(J_{\mu\alpha} p_\nu p_\beta + J_{\mu\beta} p_\alpha p_\nu \\
&+ J_{\nu\alpha} p_\mu p_\beta + J_{\nu\beta} p_\mu p_\alpha \big) +\sigma^2 (1-\sigma)^2 p_\mu p_\nu p_\alpha p_\beta \big]  
\end{align}
where $\sigma \equiv t_1 - t_2$. The above expression can be simplified greatly if one notices that any function $f(\sigma)$ that is invariant under $\sigma \to (1-\sigma)$ has the property
\begin{align}
\int_0^1 dt_1 \int_0^{t_1} dt_2 \, f(\sigma) = \frac{1}{2} \int_0^1 d\sigma \, f(\sigma)\ \ .
\end{align}
The final result then becomes 
\begin{align}\label{htrace2}
\overset{\scriptscriptstyle{(2)}}{\mathcal{H}}(s) = \frac{is^2}{2(2\pi s)^{{\text{d}}/2}} \int \frac{d^{\text{d}}p}{(2\pi)^{\text{d}}} K^{\mu\nu}_p K^{\alpha\beta}_{-p} \int_0^1 d\sigma \, e^{\sigma(1-\sigma) s p^2} \mathcal{M}_{\mu\nu\alpha\beta}
\end{align}
where
\begin{align}
\nonumber
\mathcal{M}_{\mu\nu\alpha\beta} &= \bigg[\frac{1}{4s^2}( \eta_{\mu\alpha}\eta_{\nu\beta}+ \eta_{\mu\beta}\eta_{\nu\alpha})- \frac{1}{8s} (1-2\sigma)^2 (\eta_{\mu\alpha} p_\nu p_\beta + \eta_{\mu\beta} p_\alpha p_\nu \\
&+ \eta_{\nu\alpha} p_\mu p_\beta + \eta_{\nu\beta} p_\mu p_\alpha) + \sigma^2 (1-\sigma)^2 p_\mu p_\nu p_\alpha p_\beta \bigg] \ \ .
\end{align}
All tensor structures that vanish because of the null property of the KS vector have been dropped, which comprises an extra simplification special to the KS geometry.

\subsection{Next-to-next-to-leading order}

The third diagram in figure \ref{expansion} could easily be carried out similar to the previous diagrams. Nevertheless, it has an extra subtle feature: the triangular topology of the graph with massless internal lines inevitably leads to an {\em infrared} singularity when we pass to the effective action\footnote{As long as the external legs are {\em off-shell} the singularity is soft. Yet, these singularities could disappear for specific external kinematics. For example, see the discussion in \cite{Basemtrace}.}. This is due to the {\em long-time} behavior of the heat kernel being singular. The existence of infrared singularities in gauge theory scattering amplitudes is conventionally dealt with by adding real emission graphs which guarantees all observables are IR finite \cite{Kinoshita,Lee}. Similar story takes place in gravitational scattering, see for example \cite{Weinberg,Torma}. On the contrary, the treatment of infrared singularities present in the effective action is a widely unexplored topic. It is not clear how to obtain finite predictions in this case. Although this issue is crucial for understanding non-local effective actions, its discussion lies beyond the scope of this paper. We show below that the leading non-locality is captured by the results already obtained, which suffices for the applications to be considered in this work.

\subsection{A brief comment on the result}

It is important to pause at this stage to stress that the heat kernel trace given in eqs. (\ref{htrace1}) and (\ref{htrace2}) is exact for any KS spacetime. This is indeed true regardless of the underlying gravity theory. For example, for a Schwarzschild or Kerr black hole one can use the results to obtain the {\em on-shell} one-loop effective action or the Euclidean partition function. One merely has to determine the Fourier decomposition of the KS vector and plug back in eqs. (\ref{htrace1}) and (\ref{htrace2}) in order to perform the last momentum integral. Nevertheless, we choose not to follow this pathway and present an alternative procedure which is very useful in acheiving the goals of our study. 

\section{The curvature expansion}\label{sect2}

In this section we describe in detail how to express the heat kernel trace in an expansion utilizing the geometric curvatures. Albeit being elegant, this is not the main reason why we take this direction. First, the non-local expansion is controlled by the form factors. The non-analytic logarithm in eq. (\ref{logactionintro}) is one example of a form factor. As we alluded to in the introduction, it is of utmost importance to study the covariance properties of the form factors. The first step in this direction was done in \cite{BasemQED}. An exact solution of the heat kernel over a non-trivial background spacetime supplies us with important clues about the form factors. After we display the computation, we return back to this point in section \ref{comments}. Second, having the action expanded in geometric objects facilitates the determination of the leading correction to the BH entropy. Finally, if one hopes to track the back-reaction of quantum fluctuations on the spacetime, it is desirable to express the effective action using geometric objects. 

There exist two techniques to construct the curvature expansion. The first is the {\em covariant perturbation theory} extensively developed in \cite{Barvinsky83,Barvinskyreport,Barvinsky87,Barvinsky90}. The second is {\em non-linear completion} which appeared in \cite{Basem,BasemQED}. We employ the latter which is relatively simple. The procedure here is quite similar to matching computations in effective field theories whereby the Wilson coefficients are determined. One starts by proposing a {\em local} operator basis using the classical fields and their derivatives. This basis is typically arranged as a power series expansion in {\em generalized} curvatures \cite{BasemQED}. At each order in the curvature expansion, one supplements the operators with various non-local form factors. The latter are uniquely fixed via matching onto the results obtained in the last section. We now move to apply this procedure.

\subsection{The heat kernel at zeroth order}

To zeroth order in the curvature, the only invariant available is
\begin{align}\label{hzero}
\mathcal{H}(s) = \frac{i}{(4 \pi s)^{\text{d}/2}} \int d^{\text{d}}x \, \left[\mathcal{E}_0 + \mathcal{O}(R) \right]
\end{align}
where we stripped off some factors for convenience. Here $\mathcal{E}_0$ is the form factor which will turn out to be trivial in this case. It is also important to notice that for KS spacetimes, eq. (\ref{kssimple}) holds so no factor of $\sqrt{g}$ appears. One immediately finds\footnote{This precisely comes from the first term in the expansion of eq. (\ref{traceheatmat}) where the proper-time evolution operator is approxiamted by unity.}
\begin{align}
\mathcal{E}_0 = 1 \ \ .
\end{align}

\subsection{The heat kernel at linear order}

To lowest order in the curvature, the Ricci scalar is the only invariant that can show up in the heat kernel trace
\begin{align}\label{htlinear}
\mathcal{H}(s) = \frac{i}{(4 \pi s)^{\text{d}/2}} \int d^{\text{d}}x \, \left[\mathcal{E}_0 + s\, \mathcal{G}_R(s\Box) \, R + \mathcal{O}(R^2) \right]
\end{align}
where $\Box$ is the flat space d' Alembertian and the form factor $\mathcal{G}_R(s\Box)$ can only depend on the dimensionless combination $s\Box$. The common lore in the literature is to {\em covariantize} the derivative operators but we do not adopt this approach here. More comments appear in section \ref{comments}. The matching step is most easily done in momentum space and at $\mathcal{O}(\lambda)$ the Ricci scalar reads
\begin{align}\label{Rexplinear}
\overset{\scriptscriptstyle{(1)}}{R} = \partial_\mu \partial_\nu \, K^{\mu\nu} \ \ .
\end{align}
Here the situation is subtle because the spacetime integral in eq. (\ref{htlinear}) forces the momentum variable to vanish. Hence the derivatives in the above equation forces a null result which matches the result in eq. (\ref{htrace1}). Nevertheless, we still can not determine the form factor. An alternative route is to compute the heat kernel in the {\em coincidence} limit, i.e. without invoking the spacetime integral, and then perform the matching. This way one finds a non-trivial result that enables the determination of the form factor. Let us go back to section \ref{lowestorder} and compute the coincidence limit of the heat kernel. One finds 
\begin{align}\label{heatcoin}
\overset{\scriptscriptstyle{(1)}}{H}(x,x;s) = - \frac{is}{(4 \pi s)^{\text{d}/2}} \int \frac{d^{\text{d}}p}{(2\pi)^{\text{d}}}  \, K^{\mu\nu}_p p_\mu p_\nu \int_0^1 d\sigma \, \sigma(1-\sigma) e^{\sigma(1-\sigma) sp^2} e^{-ipx}  \ \ .
\end{align}
The matching is immediate and the form factor reads
\begin{align}
\mathcal{G}_R(s\Box) = \int_0^1 d\sigma \, \sigma(1-\sigma)\, e^{-\sigma(1-\sigma) s \Box} \ \ .
\end{align} 
In appendix \ref{formfactors}, we derive a nice identity that enables us to reexpress the above result in a simpler form
\begin{align}\label{formlinear}
\mathcal{G}_R(s\Box) = \frac{1}{4} f(s\Box) + \frac{1}{2s\Box} [f(s\Box)-1]
\end{align}
where the {\em fundamental} form factor is\footnote{We stick to the name given to this special form factor in \cite{Codello}, which we find very illustrative.}
\begin{align}\label{fundff}
f(s\Box) = \int_0^1 d\sigma \, e^{-\sigma(1-\sigma)s \Box} \ \ .
\end{align}  
Later on we shall see that only the value of the form factor at zero momentum is important. In particular we find
\begin{align}
\mathcal{G}_R(0) = \frac{1}{6} \ \ .
\end{align}

\subsection{The heat kernel at quadratic order}

Along the same lines of the last section, we match the heat kernel trace given in eq. (\ref{htrace2}) onto a curvature basis. Counting the number of derivatives this must be second order in curvatures and hence
\begin{align}\label{heatquad}
\nonumber
{\mathcal{H}}(s) = \frac{i}{(4 \pi s)^{\text{d}/2}} \int d^{\text{d}}x \, &\big[\mathcal{G}_0 + s\, \mathcal{G}_R(s\Box) \, R + s^2\, R\, \mathcal{F}_R(s\Box) \, R + s^2\, R_{\mu\nu}\, \mathcal{F}_{Ric}(s\Box) \, R^{\mu\nu}  \\
&+ s^2\, R_{\mu\nu\alpha\beta}\, \mathcal{F}_{Riem}(s\Box) \, R^{\mu\nu\alpha\beta} + \mathcal{O}(R^3) \big] \ \ .
\end{align} 
We need to expand the curvature invariants to $\mathcal{O}(\lambda^2)$ which are given in appendix \ref{curvmom}. Here comes an important part of the construction: the form factor $\mathcal{G}_R(0)$ plays role in the matching procedure. Although the form factors are defined with the flat d' Alembertian the curvature tensors must be expanded appropriately. Notice as well that only $\mathcal{G}_R(0) = 1/6$ is needed as the rest of this form factor contains total derivatives and thus vanishes by momentum conservation.

Inspection of the expressions given in appendix \ref{curvmom} we see that there are three tensor structures available which appears sufficient to determine the three form factors. But in fact only two equations turn out to be independent and they read
\begin{align}
\frac{s}{48} + \frac{s^2 p^2}{8} \mathcal{F}_{Ric}(sp^2) + \frac{s^2 p^2}{2} \mathcal{F}_{Riem}(sp^2) &= \frac{s}{16} \int_0^1 d\sigma \, (1-2\sigma)^2 e^{\sigma(1-\sigma) sp^2} \label{firsteqn} \\
s^2 \mathcal{F}_{R}(sp^2) + \frac{s^2}{2} \mathcal{F}_{Ric}(sp^2) + s^2 \mathcal{F}_{Riem}(sp^2) &= \frac{s^2}{2} \int_0^1 d\sigma \, \sigma^2(1-\sigma)^2 e^{\sigma(1-\sigma) sp^2} \ \ .
\end{align}
Note the first term on the LHS of eq. (\ref{firsteqn}) which comes from $\mathcal{G}_R(0)$. We show next in detail how to uniquely fix the form factors. Once again, with the help of identities that are proven in appendix \ref{formfactors} we can express the RHS in terms of the fundamental form factor. Hence
\begin{align}
\frac{s}{48} - \frac{s^2 p^2}{8} \mathcal{F}_{Ric}(sp^2) - \frac{s^2 p^2}{2} \mathcal{F}_{Riem}(sp^2) &= \frac{1}{8 p^2} (f(sp^2)-1) \\\nonumber
\mathcal{F}_{R}(sp^2) + \frac{1}{2} \mathcal{F}_{Ric}(sp^2) + \mathcal{F}_{Riem}(sp^2) &= \frac{1}{32} f(sp^2) - \frac{1}{8sp^2} f(sp^2)\\
& + \frac{1}{16sp^2} + \frac{3}{8s^2 p^4} (f(sp^2) - 1) \ \ .
\end{align}
One might suspect that the issue we face here is special to KS spacetimes since some tensor structures vanish due to the null property of the KS vector. In fact, this is a generic feature that takes place at second order in the curvature expansion. One could easily check that the same issue arises even for an arbitrary metric, see for example \cite{Codello}.

\subsubsection{Fixing the form factors}

We saw above that there are only two available equations for three form factors that appear at second order. Usually this is circumvented by making use of the following identity \cite{Barvinsky90}
\begin{align}\label{bianchi}
\int d^4x \, \sqrt{g} \left(R_{\mu\nu\alpha\beta} (\nabla^2)^n R^{\mu\nu\alpha\beta} - 4 R_{\mu\nu} (\nabla^2)^n R^{\mu\nu} + R (\nabla^2)^n R \right) = \int d^4x \, \sqrt{g}\, \mathcal{R}^3 \ \ .
\end{align}
Here the rhs refers to cubic curvature terms. The proof of the above takes a few lines and relies on using the Bianchi identities. Hence, to second order in the curvature one can set one of the form factors in eq. (\ref{heatquad}) to zero since the {\em error} would be higher order in the curvature expansion. The canonical choice made in the literature is \cite{Barvinsky83,Barvinskyreport,Barvinsky87,Barvinsky90}
\begin{align}\label{naivechoice}
\mathcal{F}_{Riem} = 0 \ \ .
\end{align}

Indeed there is nothing special about this choice: it is nothing but one possible solution to the undetermined system of equations. Here we proceed differently because of two central reasons. First, the above choice essentially hides some of the physics contained in the computation. As we shall see below, the choice in eq. (\ref{naivechoice}) becomes dangerous when applications are considered\footnote{This point has been noted before in \cite{Basem}.}. Second, the form factors in eq. (\ref{heatquad}) strictly contain the flat space d' Alembertian and thus, formally, eq. (\ref{bianchi}) does not hold anymore.

The question remains: how can we make progress given that we have an undetermined system? This is achieved via an indirect approach, namely we consult the local UV divergences. The one-loop divergences are exactly known and expressed in a covariant manner from the coincidence limit of the Seeley-DeWitt-Gilkey series\footnote{The Seeley-DeWitt-Gilkey expansion is local and assumes a massive operator. Nevertheless, the divergences that arise at second order in the curvature are valid in the massless limit.}, see for example \cite{Birrell,Parker,Gilkey,dewitt}. Our procedure is discussed in the next section when we consider the effective action. For now we impose a seemingly {\em ad hoc} extra relation between the form factors
\begin{align}\label{extrarel}
\mathcal{F}_{Riem}(sp^2) + \mathcal{F}_{Ric}(sp^2) = 0 
\end{align}
and the consistency of this choice shall become clear in the next section. We can now solve for the form factors and find
\begin{align}\label{formquad}
\mathcal{F}_{Ric}(sp^2) &= - \mathcal{F}_{Riem}(sp^2) = \frac{1}{18sp^2} + \frac{1}{3s^2p^4} - \frac{1}{3s^2p^4} f(sp^2) \\
\mathcal{F}_{R}(sp^2) &= \frac{13}{144sp^2} - \frac{5}{24s^2p^4} + \frac{5}{24s^2p^4} f(sp^2) + \frac{1}{32} f(sp^2) - \frac{1}{8sp^2} f(sp^2) \ \ .
\end{align}
This completes the matching procedure up to this accuracy in the curvature expansion. The practice is identical if one aims to consider the $\mathcal{O}(R^3)$ basis. Nevertheless, the last diagram in figure \ref{expansion} must be computed for the matching procedure to work properly. From the vertex rules given in section \ref{sect1}, it is clear this diagram is $\mathcal{O}(\partial^6)$. Hence, the latter must be included for the non-linear completion procedure to work.

\subsection{Comments on the form factors}\label{comments}

So far we have shown how to re-express the {\em exact} results of the previous section employing the curvature expansion. One of the main concerns of the present paper is to better understand the properties of the form factors. In particular, should we enforce the following replacement?
\begin{align}\label{boxdel}
\mathcal{G}(s\Box) \to \mathcal{G}(s\nabla^2), \quad \mathcal{F}(s\Box) \to \mathcal{F}(s\nabla^2) \ \ .     
\end{align}
This is the conventional approach in the literature. Let us point out some features of the form factors that were described in \cite{BasemQED}. There - in the context of massless QED with gravitational couplings - it has been shown that the expansion of the covariant form factor $\ln(\nabla^2)$ contributes terms in the action that does {\em not} match the diagramatic expansion from perturbation theory. A proposed cure for this problem was developed in \cite{BasemQED} and referred to as the {\em counterterm method}. One has to introduce terms at higher order in the curvature expansion which are then fixed by requiring that the result matches that from perturbation theory. Albeit very complicated, it was shown that the procedure is robust and yields a unique answer for the action \cite{BasemQED}.

What does the current computation tell us about this issue? The results we presented are exact for KS spacetimes which shows that the replacement in eq. (\ref{boxdel}) is clearly superfluous. This is the main advantage of fixing the background geometry: it enables the heat kernel to be fully determined with an unambiguous definition of the form factors. Further comments appear in section \ref{sect5} regarding the fate of the form factors.

\section{The effective action}\label{sect3}

In this section, we compute the effective action up to second order in the curvature expansion. This is easily accomplished by integrating over proper time as in eq. (\ref{id}). Hence,
\begin{align}\label{EAdef}
\Gamma[g] = -\frac{i\hbar}{2} \int d^{\text{d}}x \int_0^{\infty} \frac{ds}{s} \, H(x,x;s) \ \ .
\end{align}  
 
The integral over the proper time has two interesting regimes which are known as the {\em early} and {\em late} times. The former corresponds to the small $s$ behavior and encodes the {\em short distance} behavior of the theory. The late time on the other hand corresponds to the large $s$ asymptotics of the heat kernel and controls the {\em long distance} behavior of the theory. Let us describe a simple method to uncover the UV divergences. First, recall that the heat kernel is expressed solely in terms of the fundamental form factor. We can expand the exponential in eq. (\ref{fundff}) and retain the first few terms. One then integrates over a small neighborhood, say $0 \leq s \leq 1$. The divergences then appear as a simple pole in $\epsilon$ as per usual in dimensional regularization. 

Instead of studying limits of the proper-time integral, we proceed to perform the integral all at once using a simple trick. This procedure is very useful as it reveals a close link between the UV divergences and the IR logarithmic non-locality that emerges at second order in the curvature expansion. Without any further computation, we will be able to display the answer for matter fields of various spins as well as gravitons.  

\subsection{The action at zeroth order}

If one plugs eq. (\ref{hzero}) back in eq. (\ref{EAdef}) the integral is seen to be {\em scaleless}. What should we do in this case? Let us try regulating the integral as follows
\begin{align}\label{scaleless}
\nonumber
\int_0^{\infty} ds \, s^{-\text{d}/2} &= \lim_{\delta \to 0^+}  \int_0^\infty ds \, s^{-\text{d}/2} \, e^{-\delta s} \\
&=\lim_{\delta \to 0^+} \, \delta^{\text{d}/2 - 1}\, \Gamma(1-\text{d}/2)
\end{align}
which vanishes for $\text{d} > 2$ upon taking the limit. We conclude that scaleless integrals similar to the above can be set conveniently to zero. If a mass scale was present in the operator, the above integral would yield a divergent result proportional to $m^4$, which in turn renormalizes the cosmological constant.

\subsection{The action at linear order}

We now move to the piece in eq. (\ref{htlinear}) with the form factor displayed in eq. (\ref{formlinear}). The trick to evaluate the effective action is to interchange the order of integration, namely to perform the proper time integral before the $\sigma$ integral. Once again, all scaleless integrals are dropped. We present the details of the calculation for the convenience of the reader. Let us focus on the first piece in eq. (\ref{formlinear})
\begin{align}
\nonumber
\Gamma[g] &\propto \int d^{\text{d}}x \int_0^1 d\sigma \, \int_0^\infty ds\, s^{\epsilon-2} f(s\Box)\, R \\
&= \int d^{\text{d}}x \int_0^1 d\sigma \, \int_0^\infty ds\, s^{\epsilon-2}\, e^{-\sigma (1-\sigma) s \Box} R
\end{align}
where we used $ \text{d} = 4 - 2\epsilon$. The integral over proper time is easily written in terms of the Euler gamma function
\begin{align}
\Gamma[g] \propto \int d^{\text{d}}x \int_0^1 d\sigma \, [\sigma(1-\sigma)]^{1-\epsilon}\, \Gamma(\epsilon-1)\, \Box^{1-\epsilon} R \ \ .
\end{align}  
We recognize immediately the UV divergence in the gamma function. The above expression is then expanded in $\epsilon$ and the $\sigma$ integral is readily evaluated
\begin{align}
\Gamma[g] \propto \int d^4x -\frac{1}{6} \left(\frac{1}{\bar{\epsilon}} - \ln \Box \right) \Box R, \quad \frac{1}{\bar{\epsilon}} = \frac{1}{\epsilon} - \gamma_E + \ln 4\pi
\end{align} 
where we dropped a numerical constant which amounts to a finite renormalization. The rest of the form factor is treated the same way and we end up with the divergent part
\begin{align}\label{divboxR}
\Gamma_{\epsilon}[g] = -\frac{\hbar}{60\epsilon} \int d^4x \, \Box R
\end{align}
which is indeed the correct divergence found in the Seeley-DeWitt-Gilkey expansion \cite{Birrell,Parker,Gilkey,dewitt}. In particular, dropping the scaleless integrals is fully consistent as promised. It is worth mentioning that for a massive operator, the corresponding integrals would yield divergences proportional to $m^2$ which would then renormalize Newton's constant.

More importantly is the finite IR contribution to the action which reads
\begin{align}\label{finboxR}
\Gamma[g] = \frac{\hbar}{60} \int d^4x \,  \ln \left(\frac{\Box}{\mu^2}\right) \Box R
\end{align}
where $\mu^2$ is the scale associated with dimensional regularization. Of utmost important is that the logarithmic non-locality comes tied to the UV divergence. Thus, it suffices to know the latter in order to determine the finite part of the action. It is then immediate to read off the result for any particle species other than minimally coupled scalars\footnote{We are not going to pursue this further simply because eq. (\ref{finboxR}) is not going to contribute in the applications we wish to consider. The last column in table \ref{tablecoeff} is left empty except from the scalar result that we already obtained.}. As we show below, this {\em UV-IR correspondence} continues to hold for the quadratic action. It is also crucial to point out that this correspondence is only true for the pieces in the action with four derivatives, i.e. $(\Box R, R^2)$ terms. This is easily seen by dimensional analysis: the only non-local structure that can show up is logarithmic which dictates $\log \,\mu^2$ to appear as well. The latter is a UV scale whose coefficient must be tied to the divergences. At $\mathcal{O}(R^3)$ and beyond, the one-loop effective action is finite.

\subsection{The action at quadratic order}

We now transition to the quadratic action which is the main concern of our work. The form factors are given in eq. (\ref{formquad}) and the computation proceeds similar to the previous subsection albeit one subtlty. The scaleless integrals can {\em not} be set to zero using the steps given in eq. (\ref{scaleless}): divergences are logarithmic and can not be regulated as in eq. (\ref{scaleless}). Nevertheless, let us press on by discarding those integrals as before and examine what the outcome is. Following the same steps one find the divergent piece
\begin{align}\label{divquad}
\Gamma_{\epsilon}[g] = \frac{\hbar}{32 \pi^2 \bar{\epsilon}} \int d^4x \, \left(\frac{1}{72} R^2 - \frac{1}{180} R_{\mu\nu} R^{\mu\nu} + \frac{1}{180} R_{\mu\nu\alpha\beta} R^{\mu\nu\alpha\beta} \right)
\end{align}
which is the correct set of divergences found in the Seeley-DeWitt-Glikey expansion \cite{Birrell,Parker,Gilkey,dewitt}. As advertised, dropping scaleless integrals is consistent. Here we pause to comment on the relation imposed in eq. (\ref{extrarel}). This choice was enforced based on knowledge that the divergent coefficients associated with the Riemann and Ricci pieces in eq. (\ref{divquad}) are identical but carry an opposite sign. In other words, eq. (\ref{extrarel}) is an educated guess that ensured we obtain the correct result for the effective action.

Moving on, the finite non-local portion follows immediately
\begin{align}\label{finquad}
\nonumber
\Gamma_{\text{ln}}[g] = - \frac{\hbar}{32\pi^2} \int d^4x \, &\bigg(\frac{1}{72} R \log\left(\frac{\Box}{\mu^2}\right) R - \frac{1}{180} R_{\mu\nu} \log\left(\frac{\Box}{\mu^2}\right) R^{\mu\nu} \\
&+ \frac{1}{180} R_{\mu\nu\alpha\beta} \log\left(\frac{\Box}{\mu^2}\right) R^{\mu\nu\alpha\beta} \bigg)
\end{align}
and once again we see that indeed the logarithmic non-locality is intimately tied to the divergences. This correspondence allows us to display the $\mathcal{O}(R^2)$ action given any matter field as well as gravitons from the knowledge of $\Gamma_{\epsilon}$ which is carried out below.

\subsection{The total action and renormalization}

We now carry out the renormalization program. The total action is composed of three parts 
\begin{align}\label{totalinitial}
\Gamma[\bar{g}] = \mathcal{S}_{\text{GEFT}} + \Gamma_{\epsilon} + \Gamma_{\text{ln}}
\end{align}
where $\bar{g}$ denotes the background KS metric. Here the first piece is the gravitational effective action up to $\mathcal{O}(\partial^4)$ 
\begin{align}\label{gravityeft}
\mathcal{S}_{\text{GEFT}} = \int d^4x \left( \frac{M_P^2}{2} \, R + c_1 R^2 + c_2 \, R_{\mu\nu} R^{\mu\nu} + c_3\, R_{\mu\nu\alpha\beta} R^{\mu\nu\alpha\beta} + c_4 \, \nabla^2 R\right ) \ \ .
\end{align}
Notice here that we included the Riemann tensor explicitly in the curvature basis which is not how the action is usually displayed. The last piece is usually dropped since it is a total derivative. Inspection of eq. (\ref{divboxR}) shows that we must retain this operator. Moreover, it is conventional to invoke the Gauss-Bonnet identity in order to get rid of the Riemann piece. This choice has no effect on the equations of motion. As we show in the next section, it is mandatory {\em not} to use Gauss-Bonnet in order to correctly compute the entropy. This is one crucial advantage of not adopting the naive approach - setting $\mathcal{F}_{Riem} = 0$ - as we explained in the last section. The second piece in eq. (\ref{totalinitial}) is the equivalent of eq. (\ref{divquad}) but generalized to any matter field as well as gravitons. It reads
\begin{align}\label{divspecies}
\Gamma_{\epsilon}[\bar{g}] = \frac{\hbar}{\bar{\epsilon}} \int d^4x \, \left(\alpha R^2 + \beta R_{\mu\nu} R^{\mu\nu} + \gamma R_{\mu\nu\alpha\beta} R^{\mu\nu\alpha\beta} + \Theta \Box R \right)
\end{align}
where the coefficients are listed in table \ref{tablecoeff}. Now from the UV $-$ IR correspondence uncovered before, we know how to construct the non-local portion of the action for any particle species
\begin{align}\label{logspecies}
\nonumber
\Gamma_{\text{ln}}[\bar{g}] = - \hbar \int d^4x \, &\bigg(\alpha R \log\left(\frac{\Box}{\mu^2}\right) R + \beta R_{\mu\nu} \log\left(\frac{\Box}{\mu^2}\right) R^{\mu\nu} \\
&+ \gamma R_{\mu\nu\alpha\beta} \log\left(\frac{\Box}{\mu^2}\right) R^{\mu\nu\alpha\beta} + \Theta \log\left(\frac{\Box}{\mu^2}\right) \Box R \bigg) \ \ .
\end{align} 
\begin{table}
\centering
\begin{tabular}{c|c|c|c|c}
\hline
\hline
  & $\alpha$ & $\beta$ & $\gamma$ & $\Theta$ \\
\hline
\text{Scalar} & 5 & -2 & 2 & -6 \\  
\text{Fermion} & -5 & 8 & 7 & -- \\
U(1)\text{boson} & -50 & 176 & -26 & -- \\
\text{Graviton} & 430 & -1444 & 424 & -- \\
\hline
\hline 
\end{tabular}
\caption{The coefficients appearing in the effective action due to massless fields of various spins \cite{Basem}. All numbers are divided by $11520 \pi^2$.}
\label{tablecoeff}
\end{table}

The renormalization program is now straightforward to perform by replacing the bare constants with their renormalized values\footnote{We are using the $\overline{\text{MS}}$ scheme.}
\begin{align}\label{renorm}
c_1 = c_1^r(\mu) - \frac{\alpha}{\bar{\epsilon}}, \quad c_2 = c_2^r(\mu) - \frac{\beta}{\bar{\epsilon}}, \quad c_3 = c_3^r(\mu) - \frac{\gamma}{\bar{\epsilon}}, \quad c_4 = c_4^r(\mu) - \frac{\Theta}{\bar{\epsilon}} \ \ .
\end{align}   
The renormalized constants carry an explicit scale dependence such that the renormalized action is $\mu$ independent. A standard RG analysis dictates
\begin{align}\label{RG}
\nonumber
c^r_1(\mu) &= c^r_1(\mu_\star) - \alpha \ln \left(\frac{\mu^2}{\mu_\star^2}\right) & c^r_2(\mu) &= c^r_2(\mu_\star) - \beta \ln \left(\frac{\mu^2}{\mu_\star^2}\right)\\
c^r_3(\mu) &= c^r_3(\mu_\star) - \gamma \ln \left(\frac{\mu^2}{\mu_\star^2}\right) & c^r_4(\mu) &= c^r_4(\mu_\star) - \Theta \ln \left(\frac{\mu^2}{\mu_\star^2}\right) 
\end{align}
where $\mu_\star$ is some fixed (matching) scale where the effective theory is matched onto the full theory. Clearly, the previous statement is academic since we have no knowledge of the full theory. The EFT treatment of quantum gravity is built in a bottom-up approach much like chiral perturbation theory. In such theories, the renormalized couplings must be measured experimentally \cite{JohnEFT}. When we discuss the correction to the BH entropy, we shall discover an interesting sensitivity to UV physics.

\section{The partition function and entropy}\label{sect4}

We now turn to the second goal mentioned in the introduction which is to identify the logarithmic correction to the Schwarzschild black hole entropy. On the {\em macroscopic} side, there exist a handful of methods to compute the entropy associated to a black hole. On the one hand, Gibbons and Hawking pioneered the Euclidean gravity approach \cite{Gibbons}. Subsequently, a host of Euclidean-based methods appeared in the literature as well \cite{Brown,Banados,Susskind,Solod,Garf}. On the other hand, Wald's Noether charge approach \cite{Wald1,Wald2,Wald3} expresses the entropy of a stationary black hole as an integral of a local geometric quantity - the Noether charge - over the bifurcation surface of the horizon.

One immediate advantage of knowing the effective action is to enable the use of Wald's technique. Nevertheless, the formalism as it is originally presented assumes the action to be {\em local} and a direct application of the results is not possible in our case. One general trick is to render the action local by introducing auxiliary fields and then move to apply Wald's formula. This trick was used by Myers \cite{Myers} to discuss the contribution of the Polyakov action to the entropy of 2$\text{d}$ black holes. Likewise, the authors of \cite{Aros} employed the same method to discuss the logarithmic correction to the BH entropy starting from the Riegert action \cite{Riegert}. Yet, it remains quite interesting to adapt Wald's approach to non-local field theories. We hopefully reserve this endeavor to a future publication. 

Here we choose to employ the Euclidean partition function to directly compute the entropy. Let us recall the definition of the partition function in the canonical ensemble
\begin{align}
Z(\beta) = \int_{\text{per.}} \text{D}\Psi\, \text{D}g \, e^{-\mathcal{S}_E}
\end{align}
where $\mathcal{S}_E$ is the Euclidean action, $\Psi$ denotes any matter field and $g$ is the spacetime metric. The functional integral runs over periodic field configurations, i.e. $\Psi(0,\vec{x}) = \Psi(\beta,\vec{x})$. The metrics that appear in the path-integral are those with {\em asymptotically flat} (AF) boundary conditions \cite{Gross}, i.e. approaching the flat metric on $\mathbb{R}^3 \times \mathbb{S}^1$. 

For the theory we are considering the Euclidean action reads
\begin{align}
\mathcal{S}_E = - \mathcal{S}_{\text{GEFT}} - \mathcal{S}_{\text{boundary}}  + \mathcal{S}_{\text{matter}}^E
\end{align}
where $\mathcal{S}_{\text{GEFT}}$ is given in eq. (\ref{gravityeft}), $\mathcal{S}_{\text{boundary}}$ is the Gibbons-Hawking-York boundary term \cite{Gibbons,York} and $\mathcal{S}_m^E$ is the matter action evaluated on the class of Euclidean metrics described above. Indeed one can not compute the functional integral unless some approximation is made. Note that the matter sector we consider is one-loop exact since self interactions are ignored, i.e. the path-integral is Gaussian. For metric fluctuations, we need to expand around a gravitational instanton which leads to a well-defined loop expansion for the partition function\footnote{Stationary, but non-static, black hole solutions do not have a {\em Euclidean section} \cite{Wald1}. For example, the analytic continuation of the Kerr solution yields an imaginary metric. Nevertheless, the Euclidean procedure is well-defined \cite{Gibbons}.}. At the one-loop level, the partition function now appears
\begin{align}\label{partition}
\ln Z(\beta) = \Gamma[\bar{g}_E] + \mathcal{S}_{\text{boundary}} \ \ .
\end{align}
Here, $\bar{g}_E$ is the Euclidean instanton which obeys the KS form and $\Gamma[\bar{g}_E]$ denotes the effective action evaluated {\em on-shell}.  The only subtlety here is that we have to affect the following replacement in eqs. (\ref{divspecies}) and (\ref{logspecies})
\begin{align}
\Box \to - \Delta
\end{align}
where $\Delta$  is the 4$\text{d}$ Laplacian on $\mathbb{R}^3 \times \mathbb{S}^1$.

\subsection{Schwarzschild black hole}

In this section we use the partition function to directly compute the entropy of Schwarzschild black hole. We have the fundamental relation
\begin{align}\label{entropythermal}
S = (1-\beta \partial_\beta) \ln Z(\beta) \ \ .
\end{align}

The Euclidean section of the Schwarzschild solution reads
\begin{align}
d\text{s}^2 = \left(1-\frac{2GM}{r}\right) d\tau^2 + \left(1-\frac{2GM}{r}\right)^{-1} dr^2 + r^2 \left(d\theta^2 + \sin^2\theta d\phi^2 \right)
\end{align}
with $0\leq \tau \leq \beta$. Customarily, a conical singularity at $r=2GM$ is avoided by fixing $\beta = \beta_H \equiv 8 \pi GM$ which defines the Hawking temperature. In order for us to use the effective action in eq. (\ref{totalinitial}) to evaluate the partition function, we need to affect a coordinate transformation similar to eq. (\ref{eddingtontrans}) in order to cast the above metric in its KS form. One then proceeds to carry out the spacetime integrals in eq. (\ref{totalinitial}). Although this could readily be done, the evaluation of the non-local portion in eq. (\ref{logspecies}) is quite cumbersome\footnote{The interested reader can consult \cite{Basem,Basemtrace,BasemQED} for the position-space representation of $\ln\,\Box$.}. As we show next, the logarithmic correction can be extracted in a much simpler fashion by studying the scaling properties of $\Gamma_{\text{ln}}$. 

Consider two background metrics $\bar{g}$ and $\bar{g}_\Lambda$ related as follows\footnote{One could achieve this scaling by transforming the coordinates as $x^\mu \to \Lambda x^\mu$ and simultaneously rescaling $M \to \Lambda M$.} 
\begin{align}\label{scalemetric}
\bar{g}_\Lambda = \Lambda^2\, \bar{g}
\end{align}
where $\Lambda$ is a spacetime constant. In other words, they are related by a global scale transformation. If the original metric $\bar{g}$ solves Einstein equations, so would the scaled metric. In particular, the scaled metric is an instanton. One then inquires about the corresponding change in the entropy. As evident from eq. (\ref{partition}), this requires knowledge of the transformation properties of the effective action. The various curvature tensors transform as follows
\begin{align}
\sqrt{\bar{g}_\Lambda} = \Lambda^4 \sqrt{\bar{g}}, \quad  R^\mu_{~\nu\alpha\beta}(\bar{g}_\Lambda) = R^\mu_{~\nu\alpha\beta}(\bar{g}), \quad R_{\mu\nu}(\bar{g}_\Lambda) = R_{\mu\nu}(\bar{g}), \quad R(\bar{g}_\Lambda) = \Lambda^{-2} R(\bar{g})  \ \ .
\end{align}
On the other hand, the logarithm in eq. (\ref{logspecies}) transforms as
\begin{align}
\ln\left(\frac{-\Delta}{\mu^2}\right) \to \ln\left(\frac{-\Delta}{\mu^2}\right) - \ln \Lambda^2 \ \ .
\end{align}
Finally, we have
\begin{align}\label{scaleentropy}
S_\Lambda - S \propto \ln\Lambda^2 (1-\beta \partial_\beta) \, \Upsilon[\bar{g}_E]
\end{align}
where
\begin{align}\label{upsilon}
\Upsilon[\bar{g}_E] = \int d^4x \, &\bigg(\alpha R^2 + \beta R_{\mu\nu} R^{\mu\nu} + \gamma R_{\mu\nu\alpha\beta} R^{\mu\nu\alpha\beta} - \Theta \, \Delta  R \bigg) \ \ .
\end{align}

It is easily verified that under the scale transformation in eq. (\ref{scalemetric}) the ADM mass of Schwarzschild black hole becomes
\begin{align}
M \to \Lambda M \ \ .
\end{align}
Since the mass of the black hole is the only dimensionful parameter in the solution, it is evident from eq. (\ref{scaleentropy}) that the correction to the entropy is proportional to the logarithm of the horizon area. The coefficient is easily computed from eq. (\ref{upsilon}) where only the Riemann piece contributes non-trivially. This point makes it obvious why we should keep all independent invariants present in the action\footnote{Another way to see the same physics is to realize that the Euler number of the Schwarzschild instanton is non-vanishing. Hence, a naive implementation of the Gauss-Bonnet identity is incorrect.}.

Finally, taking the local portion of the action into account we arrive at
\begin{align}\label{entropyfinal}
S_{\text{bh}} = S_{\text{BH}} + 64 \pi^2 \left(c^r_3(\mu) + \Xi \, \ln\left(\mu^2 \mathcal{A}\right) \right)
\end{align}
where $\Xi$ is given in eq. (\ref{sumparticles}). This is the second result of the paper\footnote{Notice that $\hbar$ has been set to unity.}. We observe a rather important feature in the result: the entropy is invariant under RG evolution
\begin{align}
\frac{d}{d \ln \mu} \,S_{\text{bh}} = 0 
\end{align}
where use has been made of eq. (\ref{RG}). Conversely, we could have deduced the logarithmic correction by enforcing RG invariance. Notice that $\ln \mu^2$ in eq. (\ref{logspecies}) contributes a local piece in the partition function. By dimensional consistency, there must exist a geometric quantity with the correct mass-dimension to render the logarithm dimensionless as it must be. For the Schwarzschild instanton, the only quantity available is the area of the event horizon.

\subsection{Dimensional transmutation and final remarks}\label{seccomments}

The physical character of the entropy is elegantly emphasized if we use dimensional transmutation. The constant in eq. (\ref{entropyfinal}) is dimensionless and could be traded for a dimensionful scale by writing
\begin{align} 
c^r_3(\mu) = -\, \Xi \ln \left(\mu^2 \mathcal{A}_{\text{QG}} \right) \ \ .
\end{align}
Every UV completion of quantum gravity {\em must} predict a unique value for the above constant at same matching scale. This in turn fixes the value of $\mathcal{A}_{\text{QG}}$ which has dimensions of area. In other words, the latter scale defines the theory of quantum theory. We can now rewrite eq. (\ref{entropyfinal}) with no reference to the unphysical scale $\mu$
\begin{align}
S_{\text{bh}} = S_{\text{BH}} + 64 \pi^2 \, \Xi \, \ln \left(\frac{\mathcal{A}}{\mathcal{A}_{\text{QG}}}\right)  \ \ .
\end{align}
The result exhibits a manifest correspondence between the UV and IR. This elegant dichotomy is brought about by the structure of the logarithmic non-locality in the partition function. Here, one clearly sees the power of the EFT framework. Induced by the non-analytic portion of the action, the logarithmic dependance on the horizon area and the associated coefficient furnish a test laboratory for any proposed theory of quantum gravity. Yet, a short-distance scale, characaterictic of the UV completion, shows up hand-in-hand with the infrared effect.

Some remarks are due in place. It is quite intriguing that the coefficient of the logarithm in eq. (\ref{sumparticles}) is not positive definite. The gauge fields in the theory yields a negative contribution. In fact, dialing up the number of particles could render the quantum correction large even in a regime of weak curvature (large mass). In other words, the logarithm might compete with the BH term in the large-N limit. The inevitable existence of massless gauge fields makes it possible to attain a state of vanishing entropy. Nevertheless, it is not clear to us if this observation hides any deep physics. One might also inquire if higher curvature (loop) corrections would alter the result. The uncovered UV/IR properties of the correction lead us to believe that the logarithmic correction does not receive any modification.  

\section{Future outlook}\label{sect5}

There exist a handful of open questions which we reserve for future work. Let us outline them in some detail:

\begin{itemize}
\item The fate of the form factors and their covariance properties remains unclear in an arbitrary spacetime. In our case, we lost {\em general} coordinate invariance by fixing the background geometry to be a KS spacetime. Yet, we gained the ability to obtain the exact effective action up to second order in the curvature. In particular, we uncovered the non-analytic structure of the form factors which turned out to be rather simple. Only the flat space derivative operators appear in the form factors. The counterterm method initiated in \cite{BasemQED} was {\em unnecessary} in our construction. More work is needed to clarify if there exists a better way to display the answer in a generic spacetime. 

\item To realize a successful program of infrared quantum gravity, it is crucial to understand how to handle infrared singularities in effective actions. Although the result at second order in the curvature is free of the latter, they become omnipresent at higher orders. It was found in \cite{BasemQED} that the effective action of massless QED - with gravitational coupling - could be made infrared safe if one chooses the background fields to satisfy their lowest order equations of motion\footnote{Here, we mean both the gauge and metric fields.}. Nevertheless, this procedure is neither justified nor is it guaranteed to work. Clearly, we need further insight.   

\item Wald's Noether charge approach stands out as the most elegant technique to define and compute the entropy. In particular, it endows black hole entropy with a geometric meaning. It is rather important to obtain the logarithmic correction via Wald's approach. In 2$\text{d}$, Myers \cite{Myers} has made a pioneering step to adapt Wald's technique to study the non-local Polyakov action. Nevertheless, the non-local structure in the latter comprises a massless pole, i.e. $1/ \nabla^2$, and so it is not clear how to generalize the treatment in our case. A geometric derivation is highly desirable in order to go beyond specific black holes and generalize our results.

\item It is always interesting to derive Hawking radiation using various approaches. As the effective action encodes the vacuum fluctuations, an elegant pathway to Hawking radiation should start from the effective action. Progress has been made for 2$\text{d}$ black holes, see for example \cite{Balbinot}. In 4$\text{d}$, Mukhanov et. al. \cite{Mukhanov} made an initial step in this direction by considering the contribution of {\em s-modes} to the effective action. In this case, the computation is very similar to the 2$\text{d}$ case. Nevertheless, more work needs to be done in 4$\text{d}$. 

\item Perhaps the most important future step is to study the back-reaction on the spacetime. This is mandatory in order to track the process of black hole evaporation. Much work has been devoted to study the physics in 2$\text{d}$, see for example \cite{Callan,Russo,Lowe} which is surely an incomplete list. There exist little work, if any, regarding realistic 4$\text{d}$ black holes. It is quite unlikely that one would be able to find analytic solutions to the equations of motion given the non-local structures present. Nevertheless, numerical solutions will indeed provide invaluable insight.

\end{itemize}  

\section{Acknowledgments}
I would like to thank John Donoghue and David Kastor for plenty of useful discussions and for providing feedback on the manuscript. I would like to especially acknowledge the hospitality and support of the Kavli Institute for Theoretical Physics during my stay at the Quantum Gravity Foundations: UV to IR workshop. The workshop lively environment was a particular stimulant behind this paper. This work has been supported in part by the U.S. National Science Foundation Grants No. PHY-1205896 and No. PHY11-25915.  

\appendix

\section{Kerr-Shild spacetimes}\label{KS}

For the convenience of the reader we review the derivation of the Schwarzschild solution starting from the KS ansatz for the metric. The approach presented here is due to Adler et. al. \cite{Adler}. This approach is purely algebraic which is quite different from the geometric approach originally employed by Kerr et. al. in \cite{Kerr,Schild,Debney}. 

If we substitute the metric in eq. (\ref{kssimple}) into the Ricci tensor, the vacuum Einstein equations appear as a power series in $\lambda$
\begin{align}
\sum_{i=1}^4 \overset{\scriptscriptstyle{(i)}}{R}_{\mu\nu} = 0 \ \ .
\end{align}
The expansion goes to fourth order since the Christoffel symbols truncate at second order. The Ricci tensor must vanish at each order in $\lambda$. Moreover, since $\sqrt{g}=1$, we have that $\Gamma^\mu_{\mu\nu} = 0$ and thus
\begin{align}
R_{\mu\nu} = -\partial_\alpha \Gamma^\alpha_{\mu\nu} + \Gamma^\alpha_{\beta\mu} \Gamma^\beta_{\nu\alpha}  \ \ .
\end{align}
The null property of the KS vector leads to important identities
\begin{align}
k^\mu = g^{\mu\nu} k_\nu = \eta^{\mu\nu} k_\nu, \quad k^\mu \partial_\nu k_\mu = 0 \ \ .
\end{align}
It is easy to verify that $\overset{\scriptscriptstyle{(4)}}{R}_{\mu\nu} = 0$ is satisfied. Setting $\overset{\scriptscriptstyle{(3)}}{R}_{\mu\nu} = 0$, we have another important equation
\begin{align}
\eta^{\alpha\beta} x_\alpha x_\beta = 0, \quad x_\alpha \equiv k^\beta \partial_\beta k_\alpha = k^\beta \nabla_\beta k_\alpha  \ \ .
\end{align}
Hence, $x_{\alpha}$ is null and moreover it is orthogonal to $k_\alpha$ as can easily be checked. Indeed two null vectors which are orthogonal at each point on the manifold must be proportional to each other
\begin{align}\label{geodesic}
k^\beta \nabla_\beta k_\alpha = - A k_\alpha
\end{align}
where $A$ is a scalar function\footnote{We stick to the notation of \cite{Adler} as much as possible.}. We conclude that $k^\alpha$ must be a null geodesic with non-affine parameterization. It is shown in \cite{Adler} that the $\mathcal{O}(\lambda^2)$ equation is automatically satisfied once the $\mathcal{O}(\lambda)$ equation is solved. The linear equation is elegantly expressed if we define an extra scalar function $L \equiv - \partial_\mu k^\mu$
\begin{align}\label{Olambda}
\Box(k_\mu k_\nu) = - 2 \partial_{(\mu} [(L+A) k_{\nu)}] \ \ .
\end{align}
To simplify the equations, we write the KS vector as $k_\alpha = (\kappa,\kappa \mathbf{w}) = (\kappa,\kappa w_1, \kappa w_2, \kappa w_3 )$. For stationary spacetimes, eq. (\ref{Olambda}) leads to three\footnote{The fourth equation comes from the null constraint which forces $w_i w_i =1$.} coupled second-order equations for $\kappa$ and $\mathbf{w}$. The latter could be manipulated into an equation involving only first derivatives of $\mathbf{w}$ which reads
\begin{align}
(\partial_m w_i) \, (\partial_m w_j) = P (\partial_i w_j + \partial_j w_i), \quad P \equiv \frac{L + A}{2 \kappa}  \ \ .
\end{align}

This equation is solved analytically by a linear algebraic approach \cite{Adler}. If we define a real matrix $M_{ij} \equiv \partial_i w_j$ then the above equation becomes
\begin{align}\label{basic}
M + M^T = P^{-1} M M^T \ \ .
\end{align}
Using eq. (\ref{geodesic}), one finds that $\mathbf{w}$ lies in the null space of both $M$ and $M^T$. We shall see next that the analysis is greatly simplified. Let $R$ be an orthogonal matrix defined such that
\begin{align}\label{rotation}
\mathbf{w}^\prime = R \mathbf{w}, \quad \mathbf{w}^{\prime T} = (1,0,0) \ \ .
\end{align}
Indeed the matrix $M^\prime = R^T M R$ satisfies an identical relation as eq. (\ref{basic}). Moreover, the rotated vector $\mathbf{w}^\prime$ lies in the null space of $M^\prime$. In particular, we must have 
\begin{align}
 M^\prime = \left[ \begin{array}{ccc}
0 & 0 & 0 \\
0 & N_{11} & N_{12} \\
0 & N_{21} & N_{22} \end{array} \right]
\end{align}
which yields
\begin{align}
N + N^{\prime ~ T} = P^{-1} N N^{\prime ~ T}  \ \ .
\end{align}
The above equation is easily solved in terms of an $U \in SO(2)$ matrix such that $N^\prime = P (1-U)$. The $SO(2)$ group is parameterized in terms of a single continuous variable, say $\theta$.
Plugging everything back, we find 
\begin{align}
M_{ij} = P (1-\cos \theta) (R_{2i} R_{2j} + R_{3i} R_{3j}) + P \sin \theta (R_{2i} R_{3j} - R_{3i} R_{2k}) \ \ .
\end{align}
Notice that $R$ is orthogonal and has unit determinant which enables us to write
\begin{align}
M_{ij} = P (1-\cos \theta) (\delta_{ij} - R_{1i} R_{1j}) + P \sin \theta \epsilon_{ijk} R_{1k} \ \ .
\end{align}
In particular, the elements of the first row fully determine the matrix $M$. Recall that $\mathbf{w}$ is in the null space of $M$ which forces $w_i = R_{1i}$. Finally we end up with\footnote{Notice that $\theta$ is a function of $\mathbf{w}$.}
\begin{align}\label{alphabeta}
\partial_i w_j = \alpha (\delta_{ij} - w_i \,w_j) + \beta \epsilon_{ijk} w_k, \quad \alpha \equiv P(1-\cos\theta), \quad \beta \equiv P \sin\theta  \ \ .
\end{align}
The above equation is both linear and first order in derivatives. Yet, we still need to decouple the rhs which turns out to be an exercise in vector calculus.
From the above expression we can form all possible vector and scalar quantities, i.e. $\nabla^2 \mathbf{w}$, $\nabla \cdot \mathbf{w}$ and $ \nabla \times \mathbf{w}$. Taking the triple cross product of $\mathbf{w}$ and comparing the resulting expression with $\nabla^2 \mathbf{w}$ yields an equation for the gradient of $\alpha$
\begin{align}
\nabla \alpha = \nabla \beta \times \mathbf{w} + (\beta^2 - \alpha^2) \mathbf{w} \ \ .
\end{align} 
From the above equation and using $\nabla \times \mathbf{w}$ we obtain a similar expression for $\beta$
\begin{align}
\nabla \beta = - \nabla \alpha \times \mathbf{w} - 2 \alpha\beta \mathbf{w} \ \ .
\end{align}
It is rather remarkable that we can remove $\mathbf{w}$ entirely from the above relations. In terms of the complex function $\gamma = \alpha + i\beta$, we compute
\begin{align}\label{poissoneik}
\nabla^2 \gamma = 0, \quad (\nabla \gamma)^2 = \gamma^4 \ \ .
\end{align}
The KS vector, and hence the specetime metric, is determined in terms of $\kappa$ and $\mathbf{w}$. This is easily acheived in terms of $\xi \equiv \gamma^{-1}$. A straightforward manipulation of $\nabla \xi \times \nabla \xi^\star$ and $\nabla \xi \cdot \nabla \xi^\star$ yields the desired result
\begin{align}
\mathbf{w} = \frac{i \nabla \xi \times \nabla \xi^\star + \nabla \xi + \nabla \xi^\star}{(  1 + \nabla \xi \cdot \nabla \xi^\star)} \ \ .
\end{align}
It remains to find $\kappa$. We note that eq. (\ref{Olambda}) yields 
\begin{align}
\nabla^2 (\kappa^2 \mathbf{w}) = \nabla [(L+A) \kappa], \quad \nabla^2 \kappa^2 = 0 \ \ .
\end{align}
Remarkably, these two equations are simultaneously satsified with the choice $\kappa^2 = c \,\alpha$, where $c$ is an arbitrary constant. 

Let us apply the formalism to find the Schwarzschild solution. A real function solving eq. (\ref{poissoneik}) is transparent
\begin{align}
\gamma = \frac{c}{r} = \frac{c}{(x^2+y^2+z^2)^{1/2}} \to k_\mu = \frac{c}{\sqrt{r}} \left(1,\frac{\mathbf{x}}{r} \right)
\end{align}
which yields 
\begin{align}
ds^2  = dt_\star^2 - (d\mathbf{x} \cdot d\mathbf{x}) - \frac{c^2}{r} \left(dt_\star + dr \right)^2 \ \ .
\end{align}
This is the Schwarzschild solution in Eddington coordinates. A simple coordinate transformation
\begin{align}\label{eddingtontrans}
t_\star = t + c^2 \ln(r/c^2 - 1)
\end{align}
yields the usual form of the Schwarzschild metric. The free constant is determined as per usual from the Newtonian limit of the solution, $c^2 = 2GM$.

\section{Heat Kernel}\label{hkernel}

\subsection{Definition}\label{hkdef}

At the one-loop level, one is interested in computing a functional trace of the logarithm of some operator. That is
\begin{align}\label{EA}
\Gamma[g,\Phi] \propto  \text{Tr} \ln\left(\frac{\mathcal{D}}{\mathcal{D}_0}\right)
\end{align}
where $\Phi$ comprises extra background fields present in the system and $\text{Tr}$ denotes a trace operation over spacetime as well as internal degress of freedom. Using the identity
\begin{align}\label{id}
\ln\left(\frac{\mathcal{D}}{\mathcal{D}_0}\right) = \int_0^\infty \frac{ds}{s} \left(e^{-s\mathcal{D}_0} - e^{-s\mathcal{D}}\right),
\end{align}
the heat kernel is defined as follows
\begin{align}\label{heat}
H(x,y;s) = e^{-s\mathcal{D}} \delta^{(\text{d})}(x-y) \ \ .
\end{align}
The parameter $s$ is conventionally called {\em proper time}. Notice that the Dirac-delta distribution is not covariant in the above expression\footnote{The delta distribution contains an implicit identity tensor acting in field space.}. This choice of normalization appeared in \cite{Codello} and is convenient for our purposes. The eigenmodes of the operator $\mathcal{D}$ are tensor {\em densities} of weight $1/2$ normalized as follows
\begin{align}\label{norm}
\mathcal{D} \varphi_n = \lambda_n \varphi_n, \quad \int d^{\text{d}}x \varphi_n \,\varphi_m = \delta_{nm}, \quad \delta^{(\text{d})}(x-y) = \sum_n \varphi_n(x) \varphi_n(y)  \ \ .
\end{align}
Hence eq. (\ref{heat}) becomes
\begin{align}
H(x,y;s) = \sum_n e^{-s \lambda_n} \varphi(x) \, \varphi(y)
\end{align}
which shows that the heat kernel defined as such is a {\em bi-tensor density} of weight $1/2$. 

The trace of the heat kernel is defined as
\begin{align}\label{heattrace}
\mathcal{H}(s) = \text{tr}_I \int d^{\text{d}}x\, H(x,x;s)
\end{align}
where $\text{tr}_I$ denotes a trace over internal degrees of freedom, i.e. spacetime indices, spin and so on. Now from eq. (\ref{heat}), we see that the heat kernel satisfies the following first order differential equation
\begin{align}\label{heatdiff}
(\partial_s + \mathcal{D}_x) H(x,y;s) = 0, \quad H(x,y;0) = \delta^{(\text{d})}(x-y) \ .
\end{align}
This last equation allows the perturbative expansion of the heat kernel to be developed.

\subsection{Perturbative expansion}\label{hkexpansion}

The heat kernel could be determined exactly if one knows the eigenvalues of the operator under consideration. This might be possible to obtain in few simple cases, for instance, Schwinger pair creation in constant electromagnetic field \cite{Julian}. In general one has to content with some sort of perturbative expansion which enables a systematic study of a certain problem. Here we describe in some detail the formalism first presented in \cite{Barvinsky83,Barvinskyreport,Barvinsky87,Barvinsky90} and reviewed in \cite{Codello}. Such formalism offers a non-local expansion of the heat kernel and is highly suitable for operators without a given mass scale and thus naturally lends itself to our computation. Recall the KS metric reads $g_{\mu\nu} = \eta_{\mu\nu} - \lambda\, K_{\mu\nu}$. Consequently, the operator reads\footnote{Any operator must start with the full spacetime d' Alembertian that results from the kinetic term in the action.}
\begin{align}
\mathcal{D} = \partial^2 + V
\end{align}
where $V$ is a function of $K_{\mu\nu}$ and any extra background fields present. Let us take $V=0$ and solve for the flat space heat kernel. Now eq. (\ref{heatdiff}) becomes
\begin{align}\label{flatheat}
(\partial_s + \partial_x^2) H_0(x,y;s) = 0
\end{align}
This is easily solved by going to Fourier space
\begin{align}
H_0(p,p^\prime;s) = (2\pi)^{\text{d}}\, \delta^{(\text{d})}(p+p^\prime) e^{sp^2}
\end{align}
which then yields
\begin{align}
H_0(x,y;s) = \frac{i}{(4\pi s)^{\text{d}/2}}\, \text{exp}\left[\frac{(x-y)^2}{4s}\right] \ \ .
\end{align}
It is convenient to introduce a matrix notation at this stage if we recognize the heat kernel as a matrix in position space. For instance, the flat-space heat kernel satisfies the following property 
\begin{align}\label{comp}
H_0(x,y;s+t) = \int d^{\text{d}}z \, H_0(x,z;s) H_0(z,y;t)
\end{align}
which could be written as
\begin{align}
H_0(s+t) = H_0(s) \times H_0(t) \ \ .
\end{align}
Note in particular the following identity
\begin{align}
\mathbb{1} = H_0(s) \times H_0(-s) \ \ .
\end{align}
To set up the perturbative expansion, we define a proper-time evolution operator as follows \cite{Codello}
\begin{align}\label{tevolution}
U(s) = H_0(-s) \times H(s)
\end{align}
which, using eqs. (\ref{heatdiff}) and (\ref{flatheat}), is easily seen to satisfy the following differential equation
\begin{align}\label{difftev}
\partial_s U(s) = - H_0(-s)\times V  \times H(s), \quad U(0) = \mathbb{1} \ \ .
\end{align}
Now the interaction $V$ is also a matrix in position space. The above equation is not yet in the desired form, but we can use eq. (\ref{comp}) to rewrite eq. (\ref{tevolution}) as follows
\begin{align}\label{traceheatmat}
H(s) = H_0(s)\, U(s) \ \ .
\end{align}
Hence, eq. (\ref{difftev}) becomes
\begin{align}\label{difftevf}
\partial_s U(s) = - H_0(-s)\times V \times H_0(s) \times U(s)
\end{align}
and has the familiar solution
\begin{align}\label{tevolutionfinal}
U(s) = \text{T} \, \text{exp}\left(-\int_0^s \, dt\, H_0(-t) \times V \times H_0(t) \right) \ \ .
\end{align}
Here, $\text{T}$ is the proper-time ordering operator. We observe here that the proper time plays the role of $it$ in real-time perturbation theory. It proves easier to turn the integration variables into dimensionless quantities by rescaling $t \rightarrow t/s$ \cite{Codello}
\begin{align}\label{tevfinal}
U(s) = \text{T} \, \text{exp}\left(-\int_0^1 \, dt\, H_0(-st) \times V \times H_0(st)\right) \ \ .
\end{align}
This equation is the basis of the non-local expansion of the heat kernel \cite{Barvinsky83,Barvinskyreport,Barvinsky87,Barvinsky90}. We finally plug the above formula in eq. (\ref{tevolution}) to obtain the heat kernel.

\section{Useful identities}\label{usefuliden}

\subsection{The form factors}\label{formfactors}

There are various ways to relate the form factors to the fundamental one in eq. (\ref{fundff}). Let us process the following integral
\begin{align}
I(n) = \frac{1}{2} \int_0^1 d\sigma \, (1-2\sigma)^2 \left(\sigma(1-\sigma)\right)^n
\end{align}
which is easily expressed in terms of the Euler gamma function
\begin{align}
\nonumber
I(n) &= \frac{1}{n+1} \frac{\Gamma(2+n) \Gamma(2+n)}{\Gamma(4+2n)} 
 \ \ .
\end{align}
This can be put back into an integral representation
\begin{align}
I(n) = \frac{n!}{(n+1)!} \int_0^1 d\sigma \, \sigma^{n+1} (1-\sigma)^{n+1}
\end{align}
which enables us to derive the following identity
\begin{align}
\frac{x}{2} \int_0^1 d\sigma \, (1-2\sigma)^2\, e^{\sigma(1-\sigma)x} = f(x) - 1
\end{align}
where $f(x)$ is the form factor in eq. (\ref{fundff}). Using the above, we can derive the following identities as well
\begin{align}\label{ffident}
\int_0^1 d\sigma \, \sigma(1-\sigma) e^{\sigma(1-\sigma)x} &= \frac{1}{4} f(x) - \frac{1}{2x} [f(x) - 1] \\ 
\int_0^1 d\sigma \, \sigma^2(1-\sigma)^2 e^{\sigma(1-\sigma)x} &= \frac{1}{32} f(x) - \frac{1}{8x} f(x) + \frac{1}{16x} + \frac{3}{8x^2} [f(x)-1] \ \ .
\end{align}

\subsection{Tensor integrals}\label{tensorint}

We here list the tensor integrals needed for the computation of the heat kernel.
\begin{align}
\int \, \frac{d^{\text{d}} p}{(2\pi)^{\text{d}}} \, e^{sp^2} &= \frac{i}{(4\pi s)^{\text{d}/2}} \\
\int \, \frac{d^{\text{d}} p}{(2\pi)^{\text{d}}} \, p_{\mu} p_{\nu} e^{sp^2} &= \frac{i}{(4\pi s)^{\text{d}/2}} \frac{-1}{2s} \eta_{\mu\nu} \\
\int \, \frac{d^{\text{d}} p}{(2\pi)^{\text{d}}} \, p_{\mu} p_{\nu} p_\alpha p_\beta e^{sp^2} &= \frac{i}{(4\pi s)^{\text{d}/2}} \frac{1}{4s^2} \left(\eta_{\mu\nu} \eta_{\alpha\beta} + \eta_{\mu\alpha} \eta_{\nu\beta} + \eta_{\mu\beta}\eta_{\nu\alpha}\right)
\end{align}

\subsection{Curvature invariants in momentum space}\label{curvmom}

Here we provide the momentum space representation of the different curvature invariants which are needed to determine the heat kernel at second order in the curvature. For KS spacetimes with a flat background metric in Cartesian coordinates, the quadratic invariants read at lowest order
\begin{align}
\int d^{\text{d}}x \, \text{Riem}^2 &= \frac{1}{2} \int\frac{d^{\text{d}}p}{(2\pi)^{\text{d}}} \, K_p^{\mu\nu} K_{-p}^{\alpha\beta}\, \left(p^4\, T_{\mu\nu\alpha\beta} - p^2\, \mathcal{P}_{\mu\nu\alpha\beta} + 2 p_\mu p_\nu p_\alpha p_\beta \right) \\
\int d^{\text{d}}x \, \text{Ric}^2 &= \frac{1}{8} \int\frac{d^{\text{d}}p}{(2\pi)^{\text{d}}} \, K_p^{\mu\nu} K_{-p}^{\alpha\beta} \,  \left(p^4\, T_{\mu\nu\alpha\beta} - p^2\, \mathcal{P}_{\mu\nu\alpha\beta} + 4 p_\mu p_\nu p_\alpha p_\beta \right)\\
\int d^{\text{d}}x \, R^2 &= \int\frac{d^{\text{d}}p}{(2\pi)^{\text{d}}} \, K_p^{\mu\nu} K_{-p}^{\alpha\beta} \, p_\mu p_\nu p_\alpha p_\beta
\end{align}
where we defined
\begin{align}
T_{\mu\nu\alpha\beta} = \eta_{\mu\alpha} \eta_{\nu\beta} + \eta_{\mu\beta} \eta_{\nu\alpha}, \quad \mathcal{P}_{\mu\nu\alpha\beta} = p_\mu p_\alpha \eta_{\nu\beta} + p_\mu p_\beta \eta_{\nu\alpha} + p_\nu p_\alpha \eta_{\mu\beta} + p_\nu p_\beta \eta_{\mu\alpha} \ \ .
\end{align}
We also need the expansion of the Ricci scalar to order $\lambda^2$ which reads
\begin{align}
\int\, d^{\text{d}}x \, \overset{\scriptscriptstyle{(2)}}{R} = \frac{1}{8} \int\frac{d^{\text{d}}p}{(2\pi)^d} \, K_p^{\mu\nu} K_{-p}^{\alpha\beta} \, \left(p^2 \, T_{\mu\nu\alpha\beta} - \mathcal{P}_{\mu\nu\alpha\beta} \right) \ \ .
\end{align}


\end{document}